\documentclass[floatfix,showpacs,amsmath,amssymb,letterpaper,groupaddresses,superscriptaddress]{article}
\usepackage{latexsym}
\usepackage{epsfig}

\usepackage{amsmath}
\usepackage{amssymb}
\usepackage{graphicx}
\usepackage{times}
\usepackage[section]{placeins}
\usepackage{epstopdf}

\usepackage{slashbox}

\usepackage{color}

\def\a{\alpha}
\def\b{\beta}

\def\be{\begin{equation}}
\def\ee{\end{equation}}
\def\ba{\begin{eqnarray}}
\def\ea{\end{eqnarray}}
\def\la{\langle}
\def\ra{\rangle}
\def\a{\alpha}

\def\b{\beta}

\def\h{\hskip 1cm}

\def\lo{\longrightarrow}
\def\bex{\begin{dinglist}{110}\dsquare}
	\def\eee{\end{dinglist}}
\def\bet{\begin{dinglist}{110}\bsquare}
	\definecolor{myblue}{RGB}{150,29,119}
	\definecolor{mydark}{RGB}{50,129,219}
	\newcommand{\bsquare}{\item[\color{myblue}\ding{110}]}
	\newcommand{\dsquare}{\item[\color{mydark}\ding{110}]}
\newtheorem{remark}{Remark}

\begin{document}

\vspace{4cm}
\begin{center}{\Large \bf The power of a shared singlet state, \\ compared with a  shared reference frame}\\
\vspace{2cm}
\vspace{2cm}

F. Rezazadeh,$^1$\h A. Mani,$^2$\h V. Karimipour.$^1$\\
\vspace{1cm} $^1$ Department of Physics, Sharif University of Technology, P.O. Box 11155-9161, Tehran, Iran.\\
$^2$ Department of Engineering Science, Collage of Engineering, University of Tehran, Tehran, Iran.\\

\vskip 2cm

\begin{abstract}
We show that a Shared Singlet State (SSS) can supersede a Shared Reference Frame (SRF) in certain quantum communication tasks, i.e.  tasks in which two remote players are required to estimate certain parameters of  a two-particle state sent to them.  This shows that task-specific value of resources may be somewhat  different from their common values based on  their exchange possibilities.  

\end{abstract}
\end{center}


\vskip 2cm

\section{Introduction}\label{Introduction}

Almost in any protocol of quantum communication a shared reference frame is indispensable \cite{quantum teleportation, dense coding, key distribution1, key distribution2}, otherwise correlations in measurement outcomes do not imply correlations in physical observables of remote physical systems.  
In short-distance experiments done in a single laboratory, some types of  noise like unstable fiber communication link or instability in the sending and receiving apparatus is equivalent to an unknown or varying reference frame of the receiver compared with that of the sender. To remedy this, most Quantum Key Distribution (QKD) protocols \cite{QKD3, QKD4} are supplemented by an active alignment of reference frames to eliminate the slow rotation of reference frames induced by the environment \cite{active alignment QKD1, active alignment QKD2, active alignment QKD3, active alignment QKD4, active alignment QKD5, active alignment QKD6}. Active aligning of reference frames usually is a complicated practical task and causes new problems such as lowering the secure key generation rate.
In long distance communication, i.e. between earth and satellites \cite{earth to satelite1, earth to satelite2, earth to satelite3, earth to satelite4, earth to satelite5}, atmospheric turbulence, rotation and revolution of the satellite with respect to the earth, makes it necessary to constantly align reference frames to a high precision which is again very difficult and costly.\\

On the theoretical side, precise alignment of reference frames raises basic questions:  "How much quantum resources is required for aligning a direction between two distant parties with a given precision?". This problem has been studied from different points of view
 \cite{ Gisin and Popescu, Began, Peres and Scudo, S.n, optimal measurment1, optimal measurment2, optimal measurment3, Massar}.
In particular, it was shown in \cite{Massar and Popescu} that one party (say Alice) can share a direction with another party (say Bob) by sending $N$ polarized parallel spins, where the error vanishes as $O(1/N)$. However this scaling of error is achieved only if Bob uses $N-$ particle entangled measurements which is extremely difficult from experimental point of view. In this sense we can say that establishing aligned reference frames with arbitrary precision and by using purely quantum mechanical means needs an exceedingly large amount of resources.\\
 
As an alternative method, the authors of the present article have proposed  in \cite{our paper} a method which is based on single particle measurements on entangled states shared between the two parties. In this method which is a converse of the standard QKD protocol, the players make measurements on their fixed directions and use the imperfect correlations in the publicly announced results to find the angle between their respective directions and align them accordingly. By repeating this 
procedure they can eventually fully align their coordinate systems with a precision which is as good as the method of \cite{Massar and Popescu}. \\


The problem of classical and quantum communication in the absence of Shared Reference Frames (SRF) was studied in a series of works \cite{Rudolph, relative, BRS}, where it was shown that in the absence of an SRF, it is still possible to communicate classical and quantum information if one encodes a classical bit into two and four qubits respectively. In \cite{relative} it was also shown that certain relative informations can be communicated, albeit with lower fidelity, in the absence of SRF. \\

In another development originated in \cite{Harrow} and expanded in
 \cite{vanEnk}, many other resources  like  cobit, qubit, ebit and refbit
   were introduced  and various one-way or two-way interconversion relations were proved among them. For definition of these resources see \cite{vanEnk}. The basic theme of the theory of \cite{vanEnk} is the interconversion possibility, i.e. given a finite or infinite number of resource A, is it possible or not to convert it to resource B. In this context, resource A is more valuable than resource B if A can be freely converted to B where the meaning of "freely" is determined by the constraints, i.e. local operations. In the present paper which we believe complements \cite{vanEnk} in certain sense, or makes the first steps toward such a goal,  we study the problem from another point of view, namely we focus on specific tasks and ask which resource, A or B, is more effective in performing that specific task. It may happen that a resource A is weaker than B  or  may not be comparable to it in the sense of interconversion, and yet A can perform a specific class of tasks  better than B.   \\

The tasks that we consider are of the estimation/discrimination type in which two states are sent to two remote players who do not have a shared reference frame and instead they share one singlet state. The singlet state has been shared between them by a third trusted party. We show that in these tasks, the singlet state can perform better than a shared reference frame and thus in this task-specific context, it is a more valuable resource than a shared reference frame. \\

\begin{figure}[t]
	\centering
	\includegraphics[width=15.0 cm,height=12cm,angle=0]{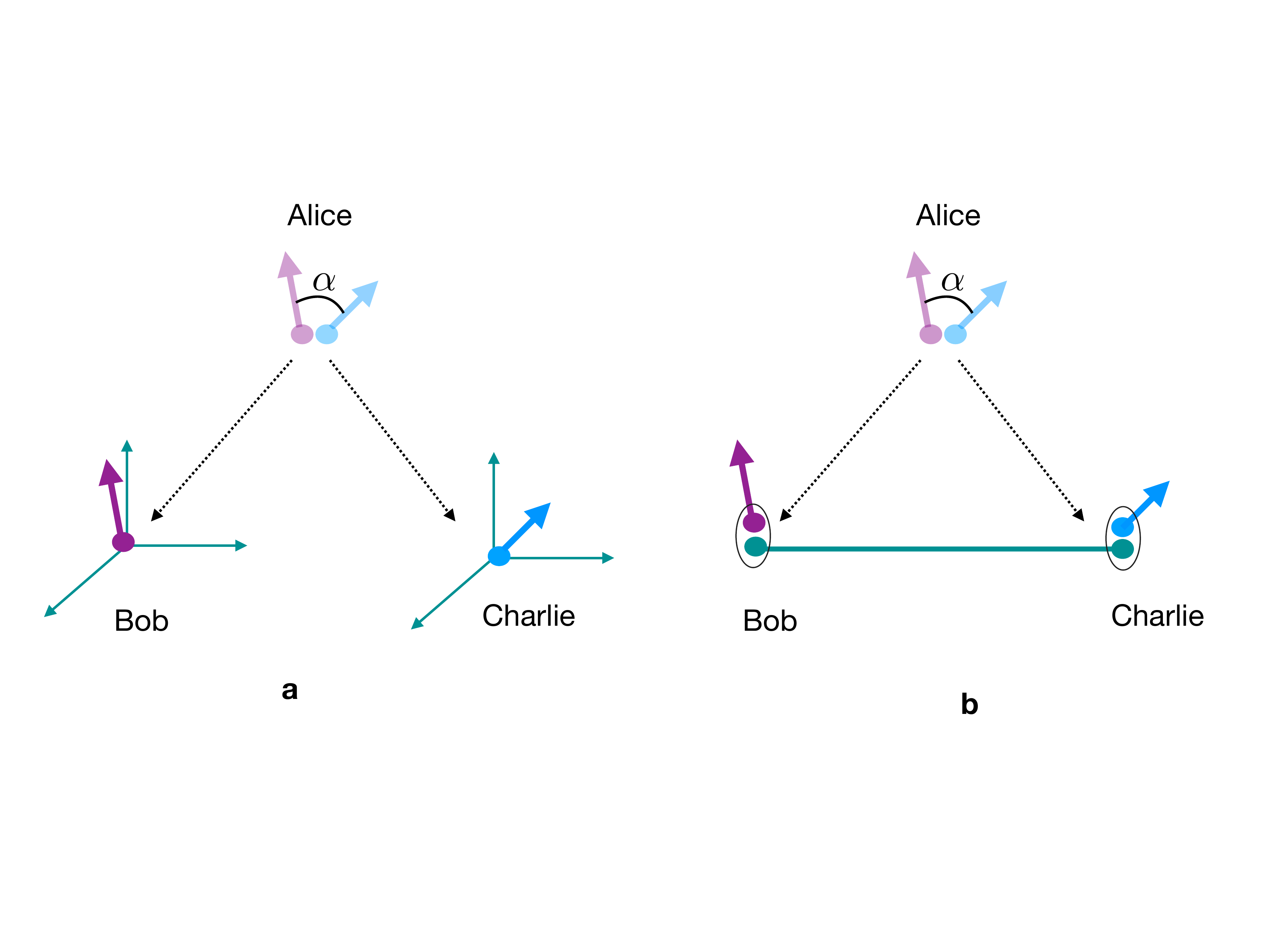}
	\vspace{-3.6 cm}
	\caption{(Color online) Task I:  Alice parallel transports two spins to Bob and Charlie who are supposed to estimate the angle between the two spins. a) when they have a shared reference frame (SRF), b) when they only have a shared singlet state (SSS), the bulbs represent total spin measurements with projectors $\Pi_0$ and $\Pi_1$. These measurements are independent of reference frame.
	}
\label{PictureB}
\end{figure}

The setup and the structure of the paper is the following: The two players are named Bob and Charlie who may either share a reference frame or a singlet state. 
Alice sends one spin 1/2 particle to Bob and another to Charlie who is far from Bob.  When they share a reference frame,  they can do single qubit measurements on the particles they receive, and when they share a singlet state,  they can do two-qubit measurement on the two particles that they hold (their share of the singlet state and the particle that they receive) figure (\ref{PictureB}). In view of the lack of any shared reference frame, the total spin measurement is the optimal measurement that each of Bob and Charlie can  perform \cite{relative}.  Three tasks are considered and compared in the forthcoming sections, namely:\\

I) Estimating the angle between the two spins, where their angle is uniformly chosen in the Alice frame, section (\ref{estimate-uniform}), \\

II) Discriminating between the spins which are parallel or anti-parallel in the Alice frame, section (\ref{estimate-pap}). \\

III) Discriminating between the case where the two spins are parallel and the case where they form a singlet in the Alice frame, section (\ref{discriminate-ps}). \\

In any case we use an appropriate measure (average information gain or probability of conclusive result) to compare the results and show that a shared singlet state is more powerful than a shared reference frame. To compare our  shared singlet state with a refbit, introduced in \cite{vanEnk}, we use refbits for performing one of the tasks, namely task II and again show the superiority of a singlet state over a refbit. We also compare the performance of singlets for various spin$-j$ singlets. Interestingly we find that when the shared singlet state is a spin$-j$ singlet, and the task is the second one, the effectiveness of this singlet state decreases with $j$ and in the limit of $j\lo \infty$, it equals that of the shared reference frame.  \\

\section{Preliminaries}

We assume that Alice prepares the  state $\rho_\a$ of two spin 1/2 particles, with a prior probability distribution $P(\a)$ which can be discrete or continuous (usually uniform). This state may or may not be a product state. She then sends one of the particles to  Bob and the other particle to Charlie who perform POVM's with elements $\{E_\lambda\}$ and $\{E'_{\lambda'}\}$ and respectively obtain the results $\lambda$ and $\lambda'$ (for short $\lambda \lambda'$) with probabilities:

\be P(\lambda \lambda'|\a):=tr\left((E_\lambda\otimes E'_{\lambda'})\rho_\a\right).\ee
They then update their knowledge of the probability distribution by using the Bayesian rule \cite{Bayesian}

\begin{equation}\label{posterior distribution}
p(\alpha\vert\lambda \lambda')=\frac{tr((E_\lambda \otimes E'_{\lambda'})\rho _\alpha)p(\alpha)}{p(\lambda  \lambda')},
\end{equation}
where

\begin{equation} \label{outprobability}
p(\lambda \lambda')=\int tr((E_\lambda\otimes E'_{\lambda'}) \rho _\alpha)p(\alpha) d\alpha.
\end{equation}
The information gain of Bob and Charlie, when they obtain the results $\lambda$ and $ \lambda'$ is given by

\begin{equation}\label{I lambda}
I_{\lambda\lambda'}=\int p(\alpha\vert\lambda\lambda') \log_2 [\frac{p(\alpha\vert\lambda\lambda')}{p(\alpha)}]d\alpha.
\end{equation}
The average information gain for all measurement results will then be

\begin{equation}\label{Iavg}
I_{avg} =\Sigma_{\lambda,\lambda'} p(\lambda\lambda') I_{\lambda\lambda'}.
\end{equation}
This is the quantity which is used for comparison of the resources in sections (\ref{estimate-uniform}) and (\ref{estimate-pap}). For the optimum measurements we use the basic result of \cite{relative} according to which in the absence of reference frames, the optimal measurements are projective measurements of the total spin for the particles of each party.
 Needless to say, in all cases Bob and Charlie have to communicate classical messages for conveying to each other the results of their measurements.   \\

\begin{remark} We have to carefully explain the bases in which we expand the states. This is specially important when Bob and Charlie do not share a reference frame, i.e. $|z_+\ra_B$ is not parallel to $|z_+\ra_C$.  Throughout the calculations we expand all the states in reference  frame of Alice and we use the rotational invariance of the singlet state to facilitate the calculations when appropriate. We also assume that the states sent by Alice to Bob and Charlie are transported through a noiseless channel, i.e. are parallel transported to Bob and Charlie.  The important point is that the singlet state can be written as $|\psi^-\ra=\frac{1}{\sqrt{2}}(|n, n^\perp\ra_{_{A,A}}-|n^\perp, n
\ra_{_{A,A}})$ for any direction $n$. Thus we always write this state in the basis of Alice, namely 
$$|\psi^-\ra=\frac{1}{\sqrt{2}}(|z_+, z_-\ra_{_{A,A}}-|z_-, z_+\ra_{_{A,A}}),$$ where $|z_-\ra_A$ and $|z_+\ra_A$ are the two basis states in the $+z$ and $-z$ direction of Alice frame. It is important to note that by rotational symmetry of the singlet, we mean $(U_A\otimes U_A)|\psi^-\ra=|\psi^-\ra$. We never use an invalid relation like  $(U_B\otimes U_C)|\psi^-\ra=|\psi^-\ra$.\\
\end{remark}
Thus all the states of this singlet and those which are sent are written in the basis $|z_+\ra_A$ and $|z_-\ra_A$, however for simplicity we do not write the subscripts in the following calculations unless there is an ambiguity which we will resolve.

\section{Task I: Estimating the angle between two spins} \label{estimate-uniform}

Alice prepares a product state of two spin $\frac{1}{2}$ particles, 
say $|\bf{n_1}\rangle \otimes|\bf{n_2}\rangle$, where $\bf{n_1}.\bf{n_2}=\cos\alpha$ and $|\bf{n_i}\rangle$ is the eigenstate of $\vec{\sigma}.\bf{n_i}$ with positive eigenvalue. She then sends one qubit to Bob and the other to Charlie. The task of Bob and Charlie is to gain the maximum possible information about the angle $\alpha $.\\

\subsection{Shared reference frame } \label{uniform-srf}

This problem has been studied in \cite{relative} where it was shown that measuring each qubit along a same (arbitrary) axis and registering whether the outcomes are the same or not is the optimal measurement for Bob and Charlie. More precisely, the implicit result of \cite{relative} is that for this specific task, the performance of one common shared direction is the same as that of a full shared coordinate system.
The maximum average information gain is then obtained to be  $0.027$ when the two spins are chosen at random on the Bloch sphere \cite{relative}, see table (\ref{relativeangletable}).

\subsection{Shared singlet state}

In this case Bob and Charlie do not share a reference frame, instead 
they share a singlet state  $|\psi^-\ra=\frac{1}{\sqrt{2}}(|z_+,z_-\ra-|z_-,z_+\ra)$ which has been sent to them by  Alice. This state is known to be rotationally invariant. Note that, as we remarked above, this state has been written in the frame of Alice. All our calculations are done in Alice frame and we never need to expand any state in the frame of Bob or Charlie which are not aligned, neither with each other nor with that of Alice. The point is that both Bob and Charlie make total spin measurements which are independent of reference frames. These measurements are denoted by POVM's $\Pi_0$ and $\Pi_1$ where the numerical subscript describes the total spin of the two particles. 
The four-qubit state provided for Bob and Charlie is
\begin{equation}\label{eqqq}
|\chi\rangle=|{\bf{n_1}}\rangle|\psi^-\rangle|{\bf{n_2}}\rangle
,
\end{equation}
in which the first two qubits are with  Bob and the third and forth ones are with Charlie. \\

When the two spins are randomly chosen from the Bloch sphere, without loss of generality we can assume that Alice selects one of them to be in a fixed direction, say $\bf{z}$ (unknown to Bob and Charlie), and the other spin randomly from a uniform distribution over the sphere. Thus $\alpha=\cos^{-1}(\bf{z} . \bf{n}_2)$ and the prior probability distribution of the  angle $\alpha$ is given by $p(\alpha)=\dfrac{1}{2}\sin(\alpha)$. Then the initial state of Bob and Charlie reads:
\begin{equation}\label{chichi}
|\chi_\alpha\rangle=|z_+\rangle(\dfrac{1}{\sqrt{2}}|z_+,z_-\rangle-\dfrac{1}{\sqrt{2}}|z_-,z_+\rangle)(\cos(\dfrac{\alpha}{2})|z_+\rangle+\sin(\dfrac{\alpha}{2})\exp(i\phi)|z_-\rangle).
\end{equation}
 where $|\psi^-\rangle$ has the same form in all bases due to its rotational invariance.\\

The elements of projective measurements of Bob and Charlie on the four qubits can be written as follows, where the first  $\Pi_i$ projects the first two particles of Bob to total spin $i$ and the second $\Pi_j$ project the last two particles with Charlie to total spin $j$:  

\begin{equation} \label{total projectors}
\Pi_{0,0}=\Pi_0\otimes\Pi_0, \h
\Pi_{0,1}=\Pi_0\otimes\Pi_1, \h
\Pi_{1,0}=\Pi_1\otimes\Pi_0, \h
\Pi_{1,1}=\Pi_1\otimes\Pi_1.
\end{equation}
Note that each of the above simple projectors like $\Pi_0$ is a two qubit measurement. \\

\begin{remark}
  {Note that neither Bob nor Charlie do not have any common reference frame with Alice and with each other. By rotationally invariant measurements of Bob and Charlie, we mean measurements on the two particles that each of them may have, one received from Alice and the other from the share of their singlet state.  The projective measurements on total spins of their two particles does not require any shared frame of reference, neither  with Alice nor with each other.}
 
 \end{remark}

The conditional probabilities of different outcomes can be calculated from $p(\Pi_{i,j}|\alpha)=\langle\chi_\a|\Pi_{i,j}|\chi_\a\rangle$. One directly calculates the matrix elements to obtain
\begin{equation}
\begin{split}
p(\Pi_{0,0}|\alpha)=\dfrac{1}{8}\sin^2(\dfrac{\alpha}{2}),\\
p(\Pi_{1,0}|\alpha)=p(\Pi_{1,0}|\alpha)=\dfrac{1}{4}-\dfrac{1}{8}\sin^2(\dfrac{\alpha}{2}),\\
p(\Pi_{1,1}|\alpha)=\dfrac{1}{2}+\dfrac{1}{8}\sin^2(\dfrac{\alpha}{2}).
\end{split}\label{PPP}
\end{equation}
The posterior distribution and average information gain depend on Bob and Charlie's prior knowledge of $\alpha$. Inserting the prior distribution  $p(\alpha)=\dfrac{1}{2}\sin(\alpha)$ in equation (\ref{posterior distribution}) will yield to the posterior probabilities
\begin{equation}
\begin{split}
P(\alpha|\Pi_{0,0})=\sin^2\dfrac{\alpha}{2}\sin\alpha, \\
P(\alpha|\Pi_{0,1})=P(\alpha|\Pi_{1,0})=(\dfrac{2}{3}-\dfrac{1}{3}\sin^2\dfrac{\alpha}{2})\sin\alpha, \\
P(\alpha|\Pi_{1,1})=(\dfrac{4}{9}-\dfrac{1}{9}\sin^2\dfrac{\alpha}{2})\sin\alpha. \\
\end{split}
\end{equation}
By inserting these probabilities in (\ref{Iavg}) and numerical integration, the average information gain will be $I_{avg}=0.0284$ which is slightly higher than the value $0.0270$ obtained with an SRF \cite{relative},  these results are summarized in table (\ref{relativeangletable}). Hence for estimating the relative parameter of two qubits, one shared singlet state is a better resource than shared reference frame. In the next two sections we will see this superiority of SSS over SRF in two other tasks.\\ 

\begin{table}[!ht]
	\centering

	\begin{tabular}{|c | c | c |} \hline
		\backslashbox{task}{resource} & SRF & SSS \\ \hline
		angle estimation & $0.0270$ & $0.0284$ \\ \hline
	\end{tabular}
	\caption{The performance of two resources for estimating the angle between two spins (quantified with the average information gain), when the spins are chosen at random on the Bloch sphere. }	\label{relativeangletable}
\end{table}

\section{Task II: Discriminating between parallel and anti-parallel spins } \label{estimate-pap}
\begin{figure}[t]
	\centering
	\includegraphics[width=15.0 cm,height=12cm,angle=0]{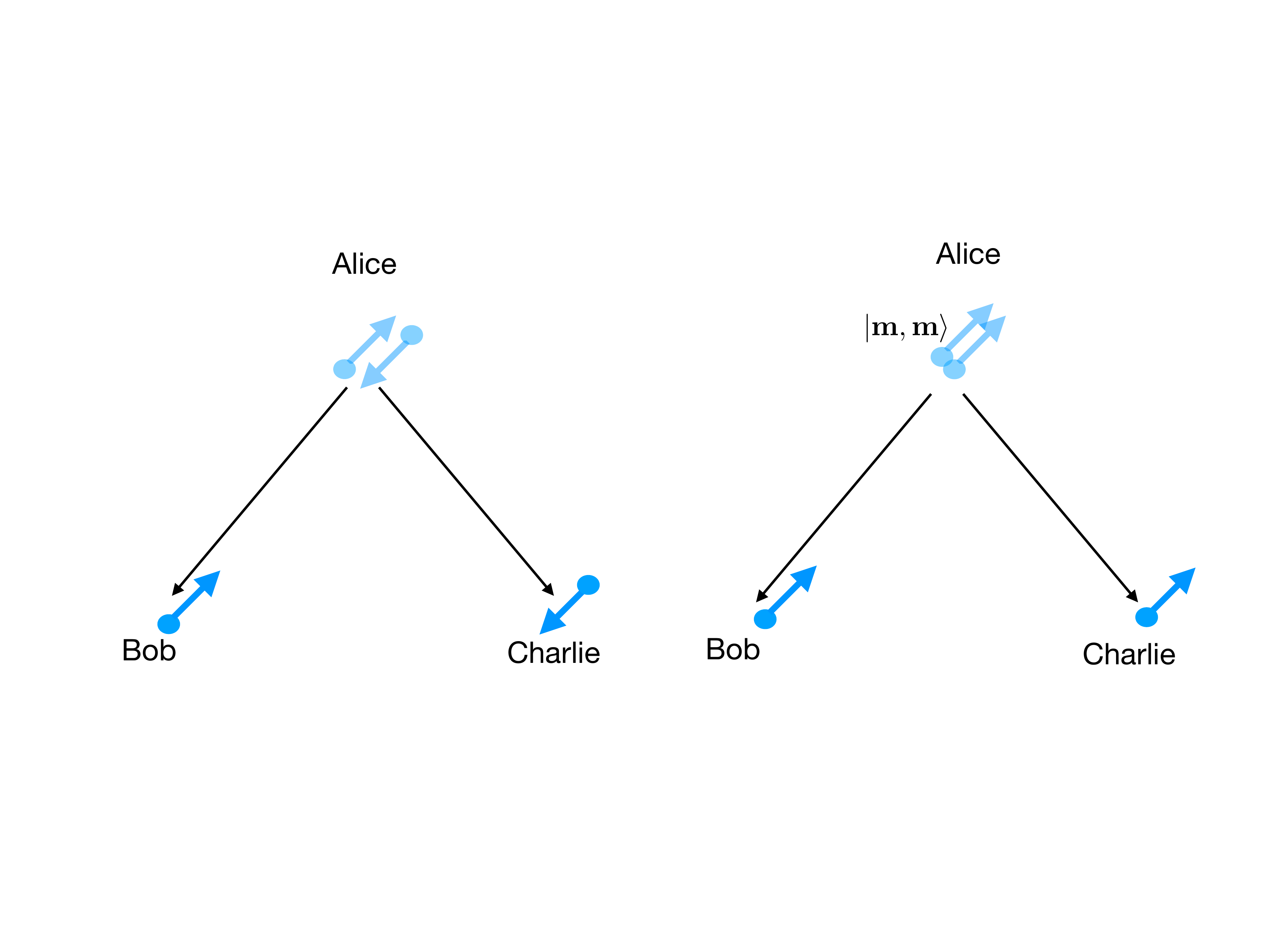}
	\vspace{-3.5 cm}
	\caption{(Color online) Task II:  Alice sends a pair of parallel or anti-parallel spins to Bob and Charlie who are to determine which pair has been sent to them.   They may  have a shared reference frame, a spin$-1/2$ shared singlet state, or a spin$-j$ singlet state. 
	}\label{picture3}
\end{figure}

Now consider the case where Alice prepares two parallel or anti-parallel spins. She then transports one spin to Bob and the other to Charlie who want to discriminate which pair is sent to them (they want to gain information about the relative angle of their received spins), see figure (\ref{picture3}). In the following subsections, we compare their performance, when they share one of the following resources, 1- A shared reference frame, 2- A spin 1/2 singlet state, 3- A higher-spin singlet state.  \\
\\


\subsection{Shared reference frame}
Again this problem has been studied in \cite{relative} and in was shown that the maximum average information gain is acquired by using the same optimal measurements explained in subsection (\ref{uniform-srf}). More explicitly, Bob and Charlie can at most gain $0.0817$ bits of information about the angle between two parallel or anti-parallel spins when they share a reference frame.

\subsection{Shared spin$-1/2$ singlet state}
In this case by inserting $p(\alpha=0)=p(\alpha=\pi)=\dfrac{1}{2}$ into (\ref{outprobability}) we will get
\begin{equation}\label{qqq}
p(\Pi_{0,0})=\dfrac{1}{16}, \h
p(\Pi_{0,1})=p(\Pi_{1,0})=\dfrac{3}{16}, \h
p(\Pi_{1,1})=\dfrac{9}{16}.
\end{equation}
Using equations (\ref{posterior distribution}) and (\ref{outprobability}) the posterior probabilities are found to be:

\begin{equation}
\begin{split}
p(\alpha=0|\Pi_{0,0})=0, \h
p(\alpha=\pi|\Pi_{0,0})=1, \\
p(\alpha=0|\Pi_{1,0})=\dfrac{2}{3}, \h
p(\alpha=\pi|\Pi_{1,0})=\dfrac{1}{3}, \\
p(\alpha=0|\Pi_{0,1})=\dfrac{2}{3}, \h
p(\alpha=\pi|\Pi_{0,1})=\dfrac{1}{3}, \\
p(\alpha=0|\Pi_{0,1})=\dfrac{4}{9}, \h
p(\alpha=\pi|\Pi_{0,1})=\dfrac{5}{9}. 
\end{split}
\end{equation}
Inserting all quantities in (\ref{I lambda}), and doing the integration, we find:
\begin{equation}
I_{\Pi_{0,0}}=1, \h
I_{\Pi_{1,0}}=I_{\Pi_{0,1}}=0.08, \h
I_{\Pi_{1,1}}=0.008.
\end{equation}

In case of obtaining $\Pi_{0,0}$, Bob and Charlie gain $1$ bit of information about the angle between the spins (since only an anti-parallel pair of spins can combine with the singlet state to produce spin zero for Both Bob and Charlie) and in the other cases much less information is acquired. The average information gained about the relative angle is found to be $$I_{avg}=\sum_{i,j}P(i,j)I_{i,j}=0.0981 \ .$$\\
This result suggests that the shared entangled state has played a role and the average information gain has  increased from $0.0817$ (for SRF) \cite{relative} to $0.0981$ (for Shared Singlet State (SSS)).\\

\subsection{The performance of higher-spin singlet states}
\begin{figure}[!ht]
	\centering
	\includegraphics[width=12.0 cm,height=8cm,angle=0]{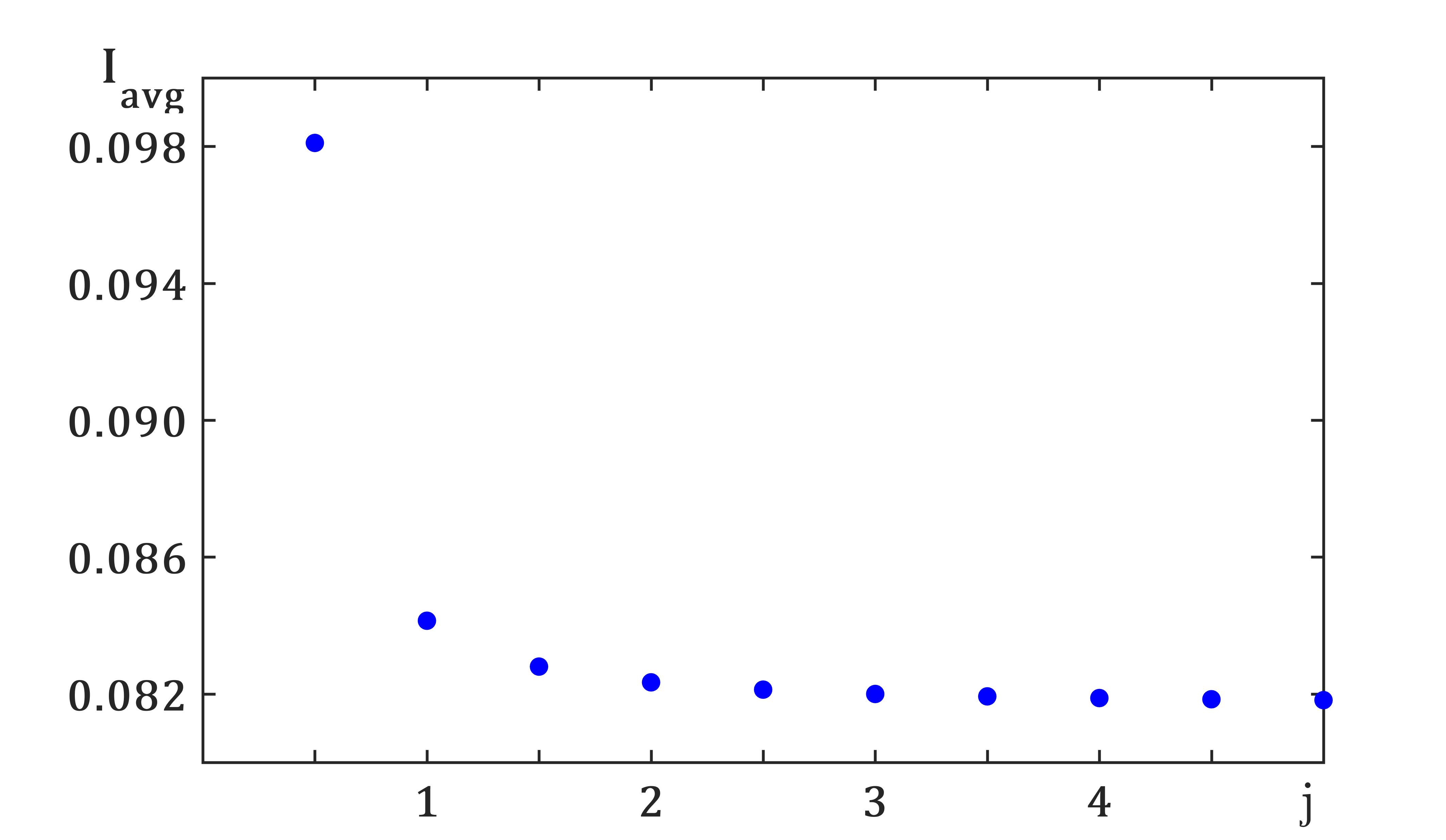}
	\vspace{0 cm}
	\caption{(Color Online) The power of a shared singlet state in performing task II decreases with spin $j$ and becomes identical to a shared reference frame for large spins. 
	}\label{spinj}
\end{figure}

The rotational invariance of the singlet state and its higher performance compared with a shared reference frame, raises a natural question: What happens if we use a higher spin singlet state, say a spin-$j$ singlet? It is intriguing to know that the performance drops with increasing $j$ and in the limit  $j\lo \infty$ it becomes equal to that of the shared reference frame.  A spin-$j$ singlet is given by 
\be
|\Psi_j^-\ra=\frac{1}{\sqrt{2j+1}}\sum_{m=-j}^j (-1)^m |j,m\ra_B| j,-m\ra_C,
\ee
where we have used the familiar angular momentum representation of states. Bob and Charlie share such a state, and their task is to get the maximum possible information about the angle between two parallel or anti-parallel spin-$\frac{1}{2}$ states which are sent to them by Alice. Again the optimal measurement is the total spin measurement on each side which is now restricted to $\Pi_{j+\frac{1}{2}}$ and $\Pi_{j-\frac{1}{2}}$. We can use the rotational invariance of the singlet state and assume that the two spins have been sent in the $z-$ basis, i.e. the parallel spins are  in the state $|z_+,z_+\ra$ and the anti-parallel spins are in the state $|z_+,z_-\ra.$  The total state which is to be locally measured by Bob and Charlie is now given by (for parallel and anti-parallel spins depending on the $+$ or $-$ sign of the last spin)
\be \label{stets-spinj}
|\Psi\ra=\frac{1}{\sqrt{2j+1}}\sum_{m=-j}^j (-1)^{m}|z_+\ra_{_{1}}|j,m\ra_{_B}\otimes |j,-m\ra_{_C}|z_\pm \ra_{_{2}},
\ee
where the states with Bob have been denoted by $1$ and $B$ and those with Charlie by $2$ and $C$.
A lengthy but straightforward calculation detailed in appendix A,  leads to the average information gain shown in figure (\ref{spinj}). It is seen that the information gain drops with increasing spin, i.e. as the singlet becomes more and more classical and in the limit of $j\lo \infty$ it becomes identical with a shared reference frame. \\

\begin{table}[!ht]
	\centering
	
	\begin{tabular}{|c | c | c | c | c |} \hline
		\backslashbox{task}{resource} & SRF & SSS \  $j=\frac{1}{2}$ & SSS \  $j=1$  & SSS \  $j \longrightarrow \infty$ \\ \hline
		Discriminating parallel and anti-parallel spins & $0.0817$ & $0.0981$ & $0.0841$ & $0.0817$ \\ \hline
	\end{tabular}
	\caption{The performance of different resources for discriminating parallel and anti-parallel spin states (quantified with the average information gain). The information gain drops with increasing spin and in the limit of $j \longrightarrow \infty$ it becomes identical with a shared reference frame. }	\label{paptable}
\end{table}

The numerical results of this section are summarized in table (\ref{paptable}). This table shows the advantage of using a shared singlet state over a shared reference frame.
In the next section we will consider another even more decisive example which shows this superiority of shared singlet state over shared frame of reference and also over a refbit \cite{vanEnk}.

\section{Task III: Discrimination between parallel spins and a singlet}\label{discriminate-ps}
\begin{figure}[t]
	\centering
	\includegraphics[width=15.0 cm,height=12cm,angle=0]{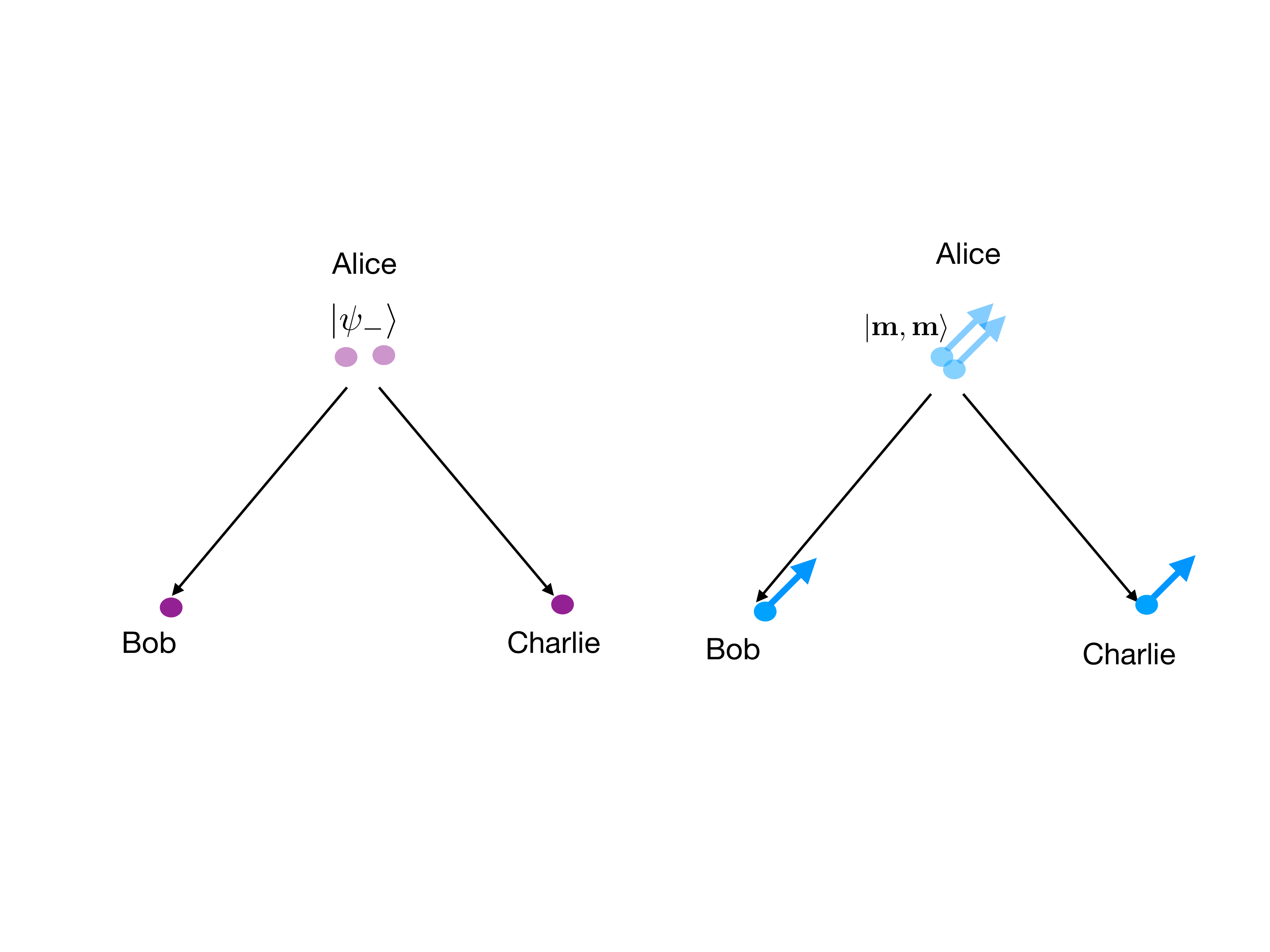}
	\vspace{- 3.6 cm}
	\caption{(Color online) Task III:  Alice randomly sends one of the two states $|\psi^-\ra$ (left) or $|{\bf m},{\bf m}\ra$ (right) to Bob and Charlie who are supposed to discriminate between the two states. They may  have a shared reference frame, a shared singlet state, or a refbit. 
	}\label{picture2}
\end{figure}

We now consider a discrimination task.  In such tasks  \cite{discrimination1,discrimination2,discrimination3} a quantum system is selected  from a known ensemble of states and sent to the participants, who are supposed to discriminate the states with minimal probability of  inconclusive results.   The ensemble is 
$$\{ |\psi^-\rangle,|{\bf m},{\bf m}\ra\}$$ where $|\psi^-\ra$ is the antisymmetric state and ${\bf m}$ is an arbitrary direction unknown to the receivers. The task of  Bob and  Charlie  is to determine which state has been sent to them, (see figure \ref{picture2}). In the following subsections, we compare their performance, when they share a shared reference frame, a shared singlet state or a refbit \cite{vanEnk}.   \\

\subsection{Shared reference frame}
 While the singlet state has the standard anti-symmetric and rotationally invariant expression, the state  $|{\bf m},{\bf m}\rangle$  can be written as
\begin{equation}\label{mm}
|{\bf m},{\bf m}\rangle=(\cos(\dfrac{\theta}{2})|z_+\rangle+\exp(i\phi)\sin(\dfrac{\theta}{2})|z_-\rangle)^{\otimes 2}.
\end{equation}
where $\theta$ and $\phi$ designate the angular coordinates of the unit vector ${\bf m}$. 
In view of the symmetry of both states with respect to exchange of particles, it is obvious that  Bob and Charlie should have the same measurement elements to obtain maximum information. So both of them measure their qubits along the same (arbitrary) axis, say $\hat{z}$. This is also in accord with the result of \cite{relative} on optimal measurements.\\

When we average over all the directions  ${\bf m}$, we find that the probability of conclusive results  is given by $\frac{1}{3}$. That is, with a shared reference frame, on the average in 1 out of 3 cases, Bob and Charlie can definitely say which of the two states $|\psi^-\ra$ or $|{\bf m}.{\bf m}\ra$ has been sent to them. 
The probabilities of each outcome $P(i,j|\psi):=tr(\Pi_{i,j}|\psi\rangle\langle\psi|)$ for $|\psi\rangle =|\psi^-\rangle $ or $|{\bf m},{\bf m}\rangle$ can be obtained by straightforward calculations. These probabilities are shown in table (\ref{table-SRF}). 
  \begin{table}[!ht] 
  	\centering
  	\begin{tabular}{|c | c | c |} \hline
  		\backslashbox{probability}{state} & $|\psi^-\rangle$ & $|{\bf m},{\bf m}\ra$ \\ \hline
  		$P(+,+|\psi)$ & 0 & $\cos^4{\theta/2}$ \\ \hline
  		$P(+,-|\psi)$ & ${1}/{2}$ & $\cos^2{\theta}/{2}\sin^2{\theta}/{2}$ \\ \hline
  		$P(-,+|\psi)$ & ${1}/{2}$ & $\cos^2{\theta}/{2}\sin^2{\theta}/{2}$ \\ \hline
  		$P(-,-|\psi)$ & $0$ & $\sin^4{\theta}/{2}$\\
  		\hline
  	\end{tabular}
  	\caption{The conditional probabilities of each outcome for different states sent by Alice when Bob and Charlie have a shared reference frame. }
  	\label{table-SRF}
  \end{table}
   
As  can be seen from this table, by obtaining the results $+,+$ or $-,-$, Bob and Charlie will be sure that Alice has sent the state $|{\bf m},{\bf m}\rangle$, while they will fail to identify the state when they  obtain $+,-$ or $-,+$. We use the probability of getting an ambiguous outcome as a figure of failure  for comparing different resources and we call it the inconclusive probability. Hence the inconclusive probability in this case is:
 \begin{eqnarray}
 \nonumber
 P_{inconclusive}&=& \dfrac{1}{2}\left[ P(+-|\psi^-)+ P(-+|\psi^-) \right]  +\ \dfrac{1}{2}\left[ P(+-|{\bf m},{\bf m})+P(-+|{\bf m},{\bf m})\right] \\
 &=& \dfrac{1}{2}+\dfrac{1}{4}\sin^2 \theta.
 \end{eqnarray}
 We consider the situation that the direction ${\bf m}$ has been chosen completely random, i.e. from a uniform distribution. The average probability of uncertainty is then found to be:
 \begin{equation} \label{uncertainSRF}
 \bar{P}_{inconclusive}= \dfrac{1}{4\pi} \int  \left( \dfrac{1}{2}+\dfrac{1}{4}\sin^2 \theta \right)  \sin\theta d\theta d\phi=\dfrac{2}{3}.
 \end{equation}
 Let us now compare this with the case where Bob and Charlie share a singlet state $|\psi^- \ra$. 

\subsection{Shared Singlet State} \label{discrimination I-SES}
 When Alice sends the state $|\psi^-\ra_{1,2}$, (when the particle 1 goes to Bob and the particle 2 goes to Charlie)  the total state of the four particles is $|\Psi\ra_{tot}=|\psi^-\ra_{B,C}\otimes |\psi^-\rangle_{1,2}$. Bob measures the total spin of the pair $(B,1)$ and Charlie measures the total spin of the pair $(C,2)$. It is obvious that the total spin of the four particles is zero and hence the outcomes $(0,1)$ and $(1,0)$ are impossible. \\

To find the probabilities of the other two outcomes, we write the total state as (note that all states are written in the Alice frame, but measured in Bob and Charlie frame as we will describe)
\be
|\Psi\ra_{tot}=\frac{1}{2}(|z_+,z_-\ra-|z_-,z_+\ra)_{B,C}\otimes (|z_+,z_-\ra-|z_-,z_+\ra)_{1,2},
\ee
which upon rearranging is given by

\be
|\Psi\ra_{tot}=\frac{1}{2}\left[|z_+,z_+\ra|z_-,z_-\ra-|z_-,z_+\ra|z_+,z_-\ra-|z_+,z_-\ra|z_-,z_+\ra+|z_-,z_-\ra|z_+,z_+\ra\right]_{B,1; C,2}.
\ee
Using the notation for the spin-1 triplet (t) and spin-0 singlet (s) and using
\ba\label{ttss}
|z_+,z_+\ra&=&|t_1\ra\cr
|z_+,z_-\ra&=&\frac{1}{\sqrt{2}}(|t_0\ra+|s_0\ra)\cr
|z_-,z_+\ra&=&\frac{1}{\sqrt{2}}(|t_0\ra-|s_0\ra)\cr
|z_-,z_-\ra&=&|t_{-1}\ra,
\ea
we find 

\be
|\Psi\ra_{tot}=\frac{1}{2}\left[|t_1\ra|t_{-1}\ra-|t_0\ra|t_0\ra+|s_0\ra|s_0\ra+|t_{-1}\ra|t_0\ra\right]_{B,1; C,2}
\ee		
Since the two parties are measuring only their total spins, this leads to $P(0,0|\psi^-)=\frac{1}{4}$ and $P(1,1|\psi^-)=\frac{3}{4}$.\\

Now suppose that Alice sends the state $|{\bf m},{\bf m}\ra_{1,2}$, so the total state of Bob and Charlie is $|\psi^-\ra_{B,C}\otimes|{\bf m}, {\bf m} \ra_{1,2}$. Bob and Charlie are going to perform the total spin measurement, due to the rotational invariance of both the singlet state and the measurements of Bob and Charlie, the results for the state $|\psi^-\ra_{B,C} \otimes |{\bf m}, {\bf m} \ra_{1,2}$ should be the same as that for $|\psi^-\ra_{B,C} \otimes |z_+, z_+ \ra_{1,2}$. Therefore we start with:
\be \label{QQ1}
|\Psi\ra_{tot}=|\psi^-\ra_{BC}\otimes |z_+,z_+\ra_{1,2}=\frac{1}{\sqrt{2}}(|z_+,z_-\ra-|z_-,z_+\ra)_{B,C}\otimes |z_+,z_+\ra_{1,2}.
\ee
Upon rearranging the labels this is written as

\ba
|\Psi_{tot}\ra_{B,1;C,2}&=&\frac{1}{\sqrt{2}} \left( |z_+,z_+\ra_{B,1} |z_-,z_+\ra_{C,2} - |z_+,z_-\ra_{B,1} |z_+,z_+\ra_{C,2} \right).
\ea
Using (\ref{ttss})  we find
\ba\label{QQ2}
|\Psi_{tot}\ra_{B,1;C,2}&=&\frac{1}{2}\left[|t_1;t_0\ra-|t_0;t_1\ra-|t_{1};s_0\ra-|s_0;t_1\ra\right]_{B1,C2}.
\ea
This equation will then lead to the probabilities shown in table (\ref{table-SES}). 
\begin{table}
	\centering
	\begin{tabular}{|c | c | c |} \hline
		\backslashbox{probability}{state} & $|\psi^-\rangle$ & $|{\bf m},{\bf m}\rangle$ \\ \hline
		$P(0,0|\psi)$ & ${1}/{4}$ & $0$ \\ \hline
		$P(0,1|\psi)$ & $0$ & ${1}/{4}$ \\\hline
		$P(1,0|\psi)$ & $0$ & ${1}/{4}$ \\\hline 
		$P(1,1|\psi)$ & ${3}/{4}$ & ${1}/{2}$ \\\hline
	\end{tabular}
	\caption{The conditional probabilities of each outcome for different states sent by Alice, when Bob and Charlie have a shared singlet state. }
	\label{table-SES}
\end{table}
The zero probabilities in the first three rows of table (\ref{table-SES}) shows that for these outcomes the two recipients can unambiguously discriminate the states. Only for the forth row they cannot reach a conclusion. Therefore we find
 
\begin{equation} \label{uncertainSES}
P_{conclusive}=1-\big[ \dfrac{1}{2}P(11|\psi^-)+\dfrac{1}{2}P(11|{\bf m},{\bf m})\big]=\dfrac{3}{8},
\end{equation}
which is definitely larger than the value of $\frac{1}{3}$ when Bob and Charlie shared a reference frame.  The results for this task are summarized in table (\ref{discriminate-table}).\\

\begin{table}[!ht]
	\centering
	\begin{tabular}{|c | c | c |c|} \hline
		resource & SRF & SSS & refbit \\ \hline
		probability of conclusive result & $\frac{1}{3}$ & $\frac{3}{8}$ & $\frac{1}{24}$ \\ 
		\hline
	\end{tabular}
	\caption{The performance of three different resources for  discrimination between a singlet state $|\psi^-\ra=\frac{1}{\sqrt{2}}(|\uparrow_{{\bf m}},\downarrow_{{\bf m}}\ra-|\downarrow_{{\bf m}},\uparrow_{{\bf m}}\ra)$ and two parallel spins $|\uparrow_{{\bf m}},\uparrow_{{\bf m}}\ra$. The performance is measured by the probability of unambiguous discrimination averaged over all input states. }\label{discriminate-table}
	\vspace{0.5cm}
\end{table}

\subsection{Comparison with Refbit}\label{reff}
One of the resources which has been introduced in the literature is called "refbit"  \cite{vanEnk} which can be considered as one unit of sharing a reference frame.  A refbit is defined as a pair of parallel spins pointing to a specific direction $|\uparrow_{{\bf n}},\uparrow_{{\bf n}}\ra$ , shared between the parties who are unaware of the direction ${\bf n}$ but are assured of them being parallel. It is obvious that an SSS can easily be transformed to a refbit by local operations. Bob simply measures his qubit in a direction of his choice ${\bf n}$ which collapses the state to a product of two anti-parallel spins, say $|\uparrow_{{\bf n}}, \downarrow_{{\bf n}}\ra$. Since the direction of ${\bf n}$ is known to Bob due to his own measurement, this  state can then be transformed into $|\downarrow_{{\bf n}},\downarrow_{{\bf n}}\ra$ by a NOT operation.  Therefore an SSS is easily transformed to pair of parallel or anti-parallel spins (i.e. a refbit). This means that it can supersede a refbit in any type of discrimination or estimation task. To make this explicit, we compare the performance of these two resources in this  task.\\


Without loss of generality we can take the refbit to be $|z_+,z_+\ra$ in the frame of Alice. 
 The four-qubit states of Bob and Charlie, when Alice sends two parallel spins $|{\bf m},{\bf m}\ra$ and a singlet, are respectively given by 
\be
|\phi_1\rangle = |z_+\rangle|\textbf{m}\rangle|\textbf{m}\rangle|z_+\rangle,
\ee
\be
|\phi_2\rangle = |z_+\rangle (\dfrac{1}{\sqrt{2}}(|z_+,z_-\rangle - |z_-,z_+\rangle))|z_+\rangle.
\ee

Bob and Charlie should perform a suitable two-qubit measurement on their own qubits and communicate their results to determine with certainty which state has been sent to them. In  appendix B we have shown that, even when they share a refbit,  the optimal measurement that minimizes the inconclusive probability is still the total spin measurement. The probabilities of each measurement outcome $P(i,j|\phi_k):=tr(\Pi_{i,j}|\phi_k\rangle\langle\phi_k|)$ for $i=0,1$ and $k=1,2$ can be obtained straightforwardly by writing $|\phi_1 \rangle$ and $|\phi_2\rangle $ in terms of the total spin eigenvectors. The results are shown in table (\ref{table-refbit}).\\

\begin{table}[!ht]
	\centering
	\begin{tabular}{|c | c | c |} \hline
		\backslashbox{probability}{state} & $|\psi^-\rangle$ & $|{\bf m},{\bf m}\rangle$ \\ \hline
		$P(0,0|\psi)$ & $0$ & ${1}/{4}Sin^4(\dfrac{\theta}{2})$ \\  \hline
		$P(0,1|\psi)$ & ${1}/{4}$ & ${1}/{4}Sin^4(\dfrac{\theta}{2})+{1}/{2}Sin^2(\dfrac{\theta}{2})Cos^2(\dfrac{\theta}{2})$ \\\hline
		$P(1,0|\psi)$ & ${1}/{4}$ & ${1}/{4}Sin^4(\dfrac{\theta}{2})+{1}/{2}Sin^2(\dfrac{\theta}{2})Cos^2(\dfrac{\theta}{2})$ \\\hline 
		$P(1,1|\psi)$ & ${1}/{2}$ & ${1}/{4}Sin^4(\dfrac{\theta}{2})+Sin^2(\dfrac{\theta}{2})Cos^2(\dfrac{\theta}{2})+Cos^4(\dfrac{\theta}{2})$ \\\hline
	\end{tabular}
	\caption{The conditional probabilities of each outcome for different states sent by Alice, when Bob and Charlie have a refbit. }
	\label{table-refbit}
\end{table}

According to table (\ref{table-refbit}), Bob and Charlie can discriminate the state unambiguously only if their measurement results is $(0,0)$. Hence the inconclusive probability is:

\begin{eqnarray}
\nonumber
P_{inconclusive}&=&1-[\dfrac{1}{2}P(0,0|\psi^-)+\dfrac{1}{2}P(0,0|{\bf m},{\bf m})]\\
&=& 1-\dfrac{1}{8}\sin^4(\dfrac{\theta}{2}).
\end{eqnarray}
The average probability of uncertainty is obtained by integrating over $\theta$:

\begin{equation} 
\bar{P}_{inconclusive}= \dfrac{1}{4\pi} \int  \left( 1-\dfrac{1}{8}\sin^4(\dfrac{\theta}{2}) \right)  \sin\theta d\theta d\phi=\dfrac{23}{24}.
\end{equation}\label{lastequation}
which is definitely larger than $\dfrac{5}{8}$ and shows the superiority of SSS over refbit. \\

It is valuable to note here that one could also use the probability of conclusive result as a figure of merit to compare the performance of SRF and SSS in task II, i.e. discrimination between parallel and anti-parallel spins. For this task, the probability of conclusive result is $\frac{1}{8}$ when the parties share a singlet state, while that is zero when they share a reference frame. This again shows the superiority of SSS over SRF for performing task II. To avoid lengthening the paper, we have presented the detailed calculations in appendix C.

 \section{Discussion}\label{conclusion}

 Usually resources are ordered according to their convertibility under the relevant constraints, i.e. local operations in entanglement theory, coherent operations in coherence theory and so on. In this sense, a resource which can be freely converted to another resource (either for finite numbers or asymptotically) is considered a stronger resource. For reference frames, such a theory has been formulated in \cite{Gour1} and \cite{Gour2}. There are situations that one can not set such ordering between different resources. i.e. some resources can not be converted to each other freely. This is exactly the case that occurs for SRF and SSS. While SSS can lead to an SRF asymptotically \cite{our paper}, an SRF can never lead to an SSS, unless it is supplied with other resources. In such cases if we consider specific tasks and ask which resource is more powerful for accomplishing that specific task, we can set an ordering. It is true that this task-specific ordering is of limited use, however in certain situations it can lead to practical advantages of a normally-considered weak resource over a strong resource. The advantage comes from the high cost for preparing the strong resource, where many of its functionalities may not be utilized for that particular task. What we have shown in this paper is an example of this kind of task-specific ordering. The ordering comes from a figure of merit which measures performance of different resources in doing specific tasks and not from convertibility of resources. Hence the constraints for different resources need not be the same, i.e. a shared singlet state is compared with a shared reference frame, more precisely a shared direction. We have also considered a refbit (a pair of parallel spins whose direction is not known to the holders) and as an example have shown that an SSS can do the discrimination task III, better than a refbit. This is of course expected since as we have shown an SSS can be freely converted to a refbit in a single shot. Finally, although we have not presented it in detail here for the sake of brevity, we have shown that not all tasks can be done better by SSS rather than an SRF. For example estimation of the angle between a spin-$j$ coherent state and a spin-1/2 state (for $j>1/2$) is an example. It is interesting to find quantum information tasks other than the ones considered in this paper, for which a Shared Singlet states (SSS) is superior to a Shared Reference Frame (SRF). It remains to be seen whether these considerations can be taken beyond a few examples and be shaped into a more general scheme for looking into resource theory of reference frames which complements the theory developed in \cite{Gour1} and \cite{Gour2}. In this paper we have restricted ourselves to one use of an SRF or SSS. For this last goal we have to make a more general study when we have multiple uses of these two resources \cite{rmk}.

\section*{Acknowledgment} The authors wish to thank L.Maccone, J.Korbicz and M.Ziman for valuable comments on this work during a poster session of IICQI-18. This work was partially supported by a grant from the vice-chancellor of Sharif University of Technology. The works of A. Mani and F. Rezazadeh was supported by a grant no. 96011347 from Iran National Science Foundation.

{}

\section*{Appendix A} \label{appendixA}
In this appendix, we briefly explain the calculations leading to figure (\ref{spinj}). Using the Clebsh-Gordon coefficients (we have used $\pm$ instead of $z_\pm$),  

\ba 
\la j, m-\frac{1}{2}; +|(j\pm \frac{1}{2}),m\ra&=&\pm \sqrt{\frac{1}{2}\big(1\pm \frac{m}{j+\frac{1}{2}}\big)}\cr
\la j, m+\frac{1}{2}; -|(j\pm \frac{1}{2}),m\ra&=& \sqrt{\frac{1}{2}\big(1\pm \frac{m}{j+\frac{1}{2}}\big)},
\ea
we rewrite (\ref{stets-spinj}) in the form
\be \label{spinj-2}
|\Psi\ra=\frac{1}{\sqrt{2j+1}}\sum_{m=-j}^j (-1)^{m}|+\ra |j,m\ra \otimes |j,-m\ra |\pm \ra 
\ee

For parallel ($\uparrow \uparrow$) and anti-parallel ($\uparrow \downarrow$) spins of Bob and Charlie, the state (\ref{spinj-2}) can be rewritten in the form

\be\begin{split}
|\Psi_{\uparrow \uparrow}\ra=\frac{1}{2j+1}\sum_{m=-j}^j(-1)^m\bigg[\sqrt{j+m+1}|j+\frac{1}{2},m+\frac{1}{2}\ra-\sqrt{j-m}|j-\frac{1}{2},m+\frac{1}{2}\ra\bigg]\otimes\\
\bigg[\sqrt{j+m+1}|j+\frac{1}{2},m+\frac{1}{2}\ra-\sqrt{j-m}|j-\frac{1}{2},m+\frac{1}{2}\ra\bigg],
\end{split}\ee
and
\be\begin{split}
	|\Psi_{\uparrow \downarrow}\ra=\frac{1}{2j+1}\sum_{m=-j}^j(-1)^m\bigg[\sqrt{j+m+1}|j+\frac{1}{2},m+\frac{1}{2}\ra-\sqrt{j-m}|j-\frac{1}{2},m+\frac{1}{2}\ra\bigg]\otimes\\
	\bigg[\sqrt{j+m}|j+\frac{1}{2},m-\frac{1}{2}\ra+\sqrt{j-m+1}|j-\frac{1}{2},m-\frac{1}{2}\ra\bigg].
\end{split}\ee\\

It is now easy to determine the probabilities of various outcomes. As an example, one finds 
\be
P(j+\frac{1}{2}\ ,\ j+\frac{1}{2}|\uparrow, \uparrow)=\frac{1}{(2j+1)^2}\sum_{m=-j}^{j}(j+m+1)^2=\dfrac{(1+j)(3+2j)}{3(1+2j)^2}.
\ee
The other probabilities are also obtained in a similar manner and we have
for parallel spins
\ba
P(j+\dfrac{1}{2} , j+\dfrac{1}{2} | \uparrow,\uparrow) &=& \dfrac{(1+j)(3+2j)}{3(1+2j)^2} \cr
P(j+\dfrac{1}{2} , j-\dfrac{1}{2} | \uparrow,\uparrow) &=& \dfrac{4j(1+j)}{3(1+2j)^2} \cr
P(j-\dfrac{1}{2} , j+\dfrac{1}{2} | \uparrow,\uparrow) &=&\dfrac{4j(1+j)}{3(1+2j)^2}, \cr
P(j-\dfrac{1}{2} , j-\dfrac{1}{2} | \uparrow,\uparrow) &=& \dfrac{j(2j-1)}{3(1+2j)^2}.
\ea
and for anti-parallel spins

\ba
P(j+\dfrac{1}{2} \otimes j+\dfrac{1}{2} | \uparrow,\downarrow) &=& \dfrac{(1+j)(3+4j)}{3(1+2j)^2}, \cr
P(j+\dfrac{1}{2} \otimes j-\dfrac{1}{2} | \uparrow,\downarrow) &=&\dfrac{2j(1+j)}{3(1+2j)^2}, \cr
P(j-\dfrac{1}{2} \otimes j+\dfrac{1}{2} | \uparrow,\downarrow) &=& \dfrac{2j(1+j)}{3(1+2j)^2}, \cr
P(j-\dfrac{1}{2} \otimes j-\dfrac{1}{2} | \uparrow,\downarrow) &=& \dfrac{j(1+4j)}{3(1+2j)^2}.
\ea
By using these probabilities one can obtain the conditional probabilities for guessing the spins to be parallel or anti-parallel and hence determine the average information gain, as in previous sections. The results are shown in figure (\ref{spinj}) which shows the average information gain versus $j$. It is seen as $j$ increases the average information gain decreases and in the limit of $j\lo \infty$, it becomes identical with the one obtained with shared reference frame. 

\section*{Appendix B}

Consider task III. Here we show that when Bob and Charlie are equipped with a refbit, the optimal measurement is the total spin measurement. Due to the lack of a complete reference frame, they can only perform rotationally invariant measurements \cite{relative}. It was shown in \cite{relative} that the elements of a rotationally invariant measurement can be expressed as a positive-weighted sum of projectors onto the total spin subspaces,i.e. 
\begin{eqnarray}\label{generalmeasure}
E_0&=&\alpha \Pi_0 + \beta \Pi_1 ,\cr
E_1&= &(1-\alpha) \Pi_0 + (1-\beta) \Pi_1.
\end{eqnarray}
where $0 \leq \alpha,\beta \leq 1$.\\

The aim is to find the best values of $\alpha$ and $\beta$ which minimize the inconclusive probability when Bob and Charlie share a refbit.
In task III Alice sends one of the states $|\psi^-\rangle$ or $|\textbf{m} , \textbf{m}\rangle$ to Bob and Charlie. After performing measurements (\ref{generalmeasure}) they have a conclusive discrimination when some of the probabilities $P(E_i,E_j|\psi^-)$ or $P(E_i,E_j|\textbf{m},\textbf{m})$ vanish and they fail to identify the state when all the probabilities are non-zero. To minimize the inconclusive probability one should maximize the probability of conclusive discrimination. Straightforward calculations shows that none of the probabilities $P(E_i,E_j|\textbf{m},\textbf{m})$ is zero and hence Bob and Charlie can have a conclusive decision if $P(E_i,E_j|\psi^-)$ vanishes for some $i$ and $j$. 
The values of $P(E_i,E_j|\psi^-)$ can be calculated easily:
\ba
P(E_0,E_0|\psi^-)&=&\dfrac{1}{2} \alpha \beta + \dfrac{1}{2} \beta^2, \cr
P(E_0,E_1|\psi^-)&=&\dfrac{1}{2} \beta (1-\beta)  + \dfrac{1}{4} \beta (1-\alpha) + \dfrac{1}{4} \alpha (1-\beta),\cr
P(E_1,E_0|\psi^-)&=&\dfrac{1}{2} \beta (1-\beta)  + \dfrac{1}{4} \beta (1-\alpha) + \dfrac{1}{4} \alpha (1-\beta),\cr
P(E_1,E_1|\psi^-)&=&\dfrac{1}{2} (1-\beta) (1-\alpha) + \dfrac{1}{2} (1-\beta)^2.
\ea
Since $0 \leq \alpha,\beta \leq 1$, it is obvious that the above probabilities can be zero only if $\beta = 0$ or $\beta = 1$, which proves the assertion. 

\section*{Appendix C}
In this appendix, by considering the probability of conclusive result as the figure of merit, we again show the superiority of SSS to SRF for performing task II (the results which we briefly explained after equation (\ref{lastequation})).
The ensemble of Alice is a pair of parallel or anti-parallel spins. Alice has sent one spin to Bob and the other to Charlie. Bob and Charlie are to determine which pair has been sent to them, see figure (\ref{picture3}). It is now better to first consider the case where Bob and Charlie share a singlet.\\
\\
\large{\textbf{C-1 \ Shared singlet state}}\\

The probabilities for the parallel spins  have already been calculated in subsection (\ref{discrimination I-SES}) and are presented in table (\ref{table-SES}). The same type of analysis as in equations (\ref{QQ1}-\ref{QQ2}) can be done for the pair of anti-parallel spins. One can rotate the reference frame of Alice so that it is aligned with the direction of this pair and so the the state of the pair is given by  $|z_+, z_-\ra$.  The calculations are straightforward and instead of (\ref{QQ1}), we now have
\ba\label{QQ11}
|\Psi_{tot}\ra_{B,1;C,2}&=&\frac{1}{\sqrt{2}}\left(|z_+,z_-\ra-|z_-,z_+\ra\right)_{B,C} \otimes |z_+,z_-\ra_{1,2}.
\ea		
which, after rearranging, in terms of total spins will be written as

\ba\label{QQ22}
|\Psi_{tot}\ra_{B,1;C,2}=\frac{1}{2\sqrt{2}}\left[ 2|t_1\ra |t_{-1}\ra - |t_0\ra |t_{0}\ra - |t_0\ra |s_0\ra 
 - |s_0\ra |t_0\ra + |s_0\ra |s_0\ra \right]_{B1;C2}.
\ea			
This will then easily leads to the probabilities shown in table (\ref{table-SES20}). 
\begin{table}
	\centering
	\begin{tabular}{|c | c | c |} \hline
		\backslashbox{probability}{state} & parallel spins & anti-parallel spins \\ \hline
		$P(0,0|\psi)$ & $0$ & $1/8$ \\ \hline
		$P(0,1|\psi)$ & $1/4$ & $1/8$ \\\hline
		$P(1,0|\psi)$ & $1/4$ & $1/8$ \\\hline 
		$P(1,1|\psi)$ & $1/2$ & $5/8$ \\\hline
	\end{tabular}
	\caption{The conditional probabilities of each outcome for two parallel or anti-parallel pairs of spins sent by Alice, when Bob and Charlie share a  singlet state. }
	\label{table-SES20}
\end{table}		
Since Alice sends her states with equal probability, it is obvious that in 1 out of 16 times, 	Bob and Charlie obtain the values $(0,0)$ which definitely lead to unambiguous discrimination. We will now see that when they have a shared reference frame, they can never reach a conclusive result.\\
\\
\large{\textbf{C-2 \ Shared reference frame}}\\

In this case, due to the arbitrariness of the direction of ${\bf m}$ and rotational invariance, the projectors of Bob can be 
$P_+:=|{\bf z}_+\ra\la {\bf z}_+|$ and $P_-=|{\bf z}_-\ra\la {\bf z}_-|$, and those of Charlie can be $P_+=|{\bf n}_+\ra\la {\bf n}_+|$ and $P_-=|{\bf n}_-\ra\la {\bf n}_-|$. To prove the inclusiveness of all measurements, it is enough to prove that none of the probabilities $P(\a,\beta| {\bf m},{\bf m})$
or $P(\a,\b| {\bf m},{\bf m}^\perp)$ ($\a,\b=\pm$))
can be zero \textit{for all choices of ${\bf m}$}. But this is an obvious fact, once we note the factorized form of the above probabilities, i.e. 
$
P(\a,\b| {\bf m},{\bf m})=|\la {\bf z}_{\pm}|{\bf m}\ra|^2|\la {\bf n}_{\pm}|{\bf m}\ra|^2
$.\\
\\
Once again we have provided that a shared singlet state is a more effective resource than a shared reference frame for performing task II. 

%
%

\end{document}